\documentstyle[12pt,epsf,epsfig]{article}
\begin{document}

\title{Understanding the Possible Proton Antiproton Bound State
Observed by BES Collaboration}
\author{Chong-Shou Gao and Shi-Lin Zhu\\
Department of Physics, Peking University, Beijing 100871, China}
\maketitle

\begin{abstract}
We comment on the quantum numbers and decay channels of the
possible proton antiproton bound state observed by BES
Collaboration. Based on the general symmetry consideration and
available experimental information, we conclude that the quantum
number of this state is very likely to be $J^{PC}=0^{-+},
I^G=0^+$, which can not decay into final states $\pi^+\pi^-,
2\pi^0, K{\bar K}, 3\pi$. Besides its dissociation into $p\bar p$,
the other important mesonic decay modes could be $\eta \pi\pi$,
$\eta'\pi\pi$, $\eta\eta\eta$, $4\pi$, $K{\bar K}\pi$, $\eta
K{\bar K}$, $K\bar K \pi\pi$, 6$\pi$. Experimental search of this
signal in these meson final states is strongly called for.

\medskip
{\large PACS number: 11.30.Er, 12.39.Mk, 13.25.Jx}
\end{abstract}
\vspace{0.3cm}

\pagenumbering{arabic}

Recently BES Collaboration in Beijing announced that they observed
a narrow enhancement near the threshold in the invariant mass
spectrum of $p \bar p$ pairs from $J/\psi \to \gamma p \bar p$
radiative decays. No similar structure is found in the $\pi^0
p\bar p$ channel \cite{bes}.

If the near threshold enhancement is fit with S-wave Breit-Wigner
resonance function, the resulting peak mass is
$M=1859^{+3}_{-10}(\mbox{stat})^{+5}_{-25}(\mbox{sys})$ MeV below
$2m_p$. The total width is less than $30$ MeV. Clearly such a mass
and width does not match that of any known particle in PDG
\cite{pdg}. With a P-wave fit, the peak mass is very close to the
threshold, $M=1876.4\pm 0.9$ MeV and the total width is very
narrow, $\Gamma =4.6\pm 1.8$ MeV. D-wave fit failed badly
\cite{bes}. The photon polar angle distribution is consistent with
$1+\cos^2 \theta_\gamma$ which suggests the angular momentum is
very likely to be $J=0$.

The study of the possible nucleon anti-nucleon bound state has a
very long history. An extensive and excellent review is given in
Ref. \cite{pr}. Interested readers are referred to Refs.
\cite{pr,history} for the general status of the experiments and
various theoretical approaches using quark model or conventional
nucleon potentials. Datta and O'Donnell described the narrow
resonance in the decay $J/\psi \to \gamma p \bar{p}$ deuteron-like
singlet state in a simple potential model with a $\lambda \cdot
\lambda$ confining interaction \cite{can}. Very recently Rosner
discussed the nature of the low-mass baryon-antibaryon enhancement
observed in B meson decays \cite{rosner}.

In this short note we will try to understand the quantum number of
this state making full use of general symmetry requirement and the
available experimental information. We will also discuss the
experimental search of this state in the possible mesonic decay
channels.

Both Quantum Chromodynamics (QCD) and Quantum Electrodynamics
(QED) conserve the charge conjugation parity. Both photon and
$J/\Psi$ have $J^{PC}=1^{--}$. So $C$-parity of this state is
positive $C=+$. For the particle and antiparticle system, we know
from textbooks that $C=(-)^{L+S}$. Positive $C$-parity requires
$L+S=2m$ where $m$ is an integer. The spin of proton is ${1\over
2}$. So the total spin of this state is $S=0, 1$.  The parity is
$P=(-)^{L+1}$. BES measurement suggests that the total angular
momentum is $J=0$ or $J=1$. When $L=S=0$, this state has
$J^{PC}=0^{-+}$. The combination of $L=2, S=0$ will lead to $J=2$
which is already excluded by BES measurement since a D-wave fit
yields unacceptable $\chi^2$. When $L=S=1$, the angular momentum
could be $J=0, 1, 2$. So, in principle, the quantum number of this
state can be $J^{PC}=1^{++}, 0^{++}$ with $L=S=1$ or
$J^{PC}=0^{-+}$ with $L=S=0$. BES's measurement of photon polar
angle distribution favors $J=0$. From now on, we mainly discuss
the possibility of $J^{PC}=0^{-+}, 0^{++}$.

Now let us move on to the isospin and $G$-parity. The production
mechanism of possible $p\bar p$ bound state is not unique in the
radiative decay $J/\Psi\to \gamma p\bar p$. For example the photon
can be produced when charm quark pairs annihilate in the reaction
$J/\Psi\to n g +\gamma \to 3 q + 3\bar q +\gamma \to p \bar p
+\gamma$ where $g$ is the virtual gluon, $n=2, 4, \cdots$ is the
number of virtual gluons and $q ({\bar q })$ is the quark
(antiquark) field. The photon was emitted from the internal charm
quark lines. The $n$ gluons hadronize into three quark and
anti-quark pairs. So the isospin of the resulting $p\bar p$ system
should be zero. If perturbative QCD is still applicable in this
energy scale, the leading order cross section is $\sim {\cal
O}(\alpha_s^4 \alpha)$ where $\alpha_s$ and $\alpha$ is the strong
and electromagnetic coupling constants respectively.

The other production mechanism is as follows : $J/\Psi \to n g \to
3 q +3\bar q\to 3 q +3\bar q +\gamma\to p \bar p +\gamma$, where
$n=3, 5, \cdots$. The $J/\Psi$ first annihilates into three or
more virtual gluons which turn into three up/down quark and
anti-quark pairs. Note the isospin of the system of the three
pairs of light quark and anti-quark is zero. Then a real photon is
emitted from one of the up or down quarks. QED violates isospin
symmetry explicitly. The isospin of the photon can be either one
or zero. After hadronization, the isospin of the final $p {\bar
p}$ system can be either one or zero. The leading order cross
section is $\sim {\cal O}(\alpha_s^6 \alpha)$, which is suppressed
by $\alpha_s^2$ compared with that of the first process.

Right now we know very little of the dynamics of this $p\bar p$
state. In the extreme case, it could be a baryonium or molecule
state of proton and antiproton bound by long-range strong force or
by electromagnetic force or by a combination of these two forces.
Or it could be a six-quark state composed of three quarks and
three antiquarks. We believe that this $p\bar p$ state is an
isoscalar to a very good accuracy. However, we also discuss the
possibility of it being an isovector for the sake of completeness.
Later we will show such a possibility is already excluded by very
recent BES analysis. Let us assume this state can have $I^G=0^+$
or $1^-$ first.

Now we have four possibilities of the quantum number of this
state: $J^{PC}I^G=0^{++}1^-$, $0^{++}0^+$, $0^{-+}1^-$,
$0^{-+}0^+$. We shall discuss the possible mesonic decay channels
of this state according to different quantum numbers.

For the strong decays of this state, the conservation of angular
momentum, parity, $C$-parity, isospin, $G$-parity, charge and four
momentum constrains the possible final states greatly even if we
know nothing about how this state is bound. If its quantum number
is $J^{PC}I^G=0^{++}1^-$, the symmetry consideration allows the
following possible meson final states \cite{gao}: $\pi^0\eta$,
$\pi^0\eta'$, $K\bar K$, $2\pi (K\bar K)$, $\eta\pi (K\bar K)$,
$\eta\eta (K\bar K)$, $5\pi$ etc. We limit ourselves to final
states less than six particles. Final states with particle number
greater than five will be too challenging for BES detector. The
dominant mesonic decay modes are possibly $K\bar K$, $\pi^0\eta$,
$\pi^0\eta'$.

With $J^{PC}I^G=0^{++}0^+$, the possible meson final states are:
$\pi\pi$, $K\bar K$, $\eta\eta$, $\eta\eta'$, $\eta'\eta'$,
$4\pi$, $\eta\pi (K\bar K)$, $\eta\eta (K\bar K)$, $2\pi (K\bar
K)$ etc. The dominant mesonic decay modes are possibly $\pi\pi,
K\bar K, 4\pi$.

With $J^{PC}I^G=0^{-+}1^-$, the possible meson final states are:
$3\pi$, $\rho\pi$, $f_0\pi$, $f_0(1370)\pi$, $a_0\eta$,
$f_0(1500)\pi$, $\eta\eta\pi$, $\eta\eta'\pi$ etc. The dominant
mesonic decay mode is possibly $3\pi$.

Very recently BES Collaboration have performed careful search of
this state in the radiative decay channels other than $\gamma p
\bar p$. No evidence of a possible bound state near the threshold
is found in the two pions, three pions, and two kaons final states
in the decays $J/\Psi\to \gamma 2\pi$, $J/\Psi\to \gamma 3\pi$ and
$J/\Psi\to \gamma K\bar K$ \cite{bes1}.

We believe that BES detector should be able to observe this
possible $p\bar p$ bound state through the radiative two-pion,
three-pion and two-kaon decay channels if its quantum number were
$J^{PC}I^G=0^{++}1^-, 0^{++}0^+, 0^{-+}1^-$. Since BES has not
seen a clear signal in the $2\pi$, $3\pi$, and $K\bar K$ decay
channels, we conclude that the quantum number of the possible
bound state observed by BES Collaboration is very likely to be
$J^{PC}I^G=0^{-+}0^+$. Such a quantum number is consistent with
BES's S-wave fit.

From now on we focus on the decay patterns of possible $p\bar p$
bound state with $J^{PC}I^G=0^{-+}0^+$. Parity and angular
momentum conservation forbids any state with such a quantum number
decaying into two pions and two kaons. $G$-parity conservation
forbids it decaying into three pions. The smaller the total number
of the necessary partial waves ${\cal L}= \Sigma l_i$, the bigger
the partial decay width as one naively expects because of phase
space suppression. We list below the possible final states
according to the total partial wave number ${\cal L}$. For ${\cal
L}=0$, the possible decay modes are $K^+ K^-\pi^0$, $K_SK_S\pi^0$,
$K_LK_L\pi^0$, $K^+{\bar {K^0}}\pi^-$, $K^-K^0\pi^+$, $\eta
K^+K^-$, $\eta K_SK_S$, $\eta K_LK_L$, $\eta\pi^+\pi^-$,
$\eta\pi^0\pi^0$, $\eta' \pi^+\pi^-$, $\eta' \pi^0\pi^0$,
$\eta\eta\eta$ \cite{gao}.

For ${\cal L}=3$, the possible decay modes are $K^+K^-\pi^+\pi^-$,
$K_LK_S\pi^+\pi^-$, $K^+ {\bar {K^0}} \pi^-\pi^0$,
$K^-K^0\pi^+\pi^0$. For ${\cal L}=5$, the possible decay modes are
$\pi^+\pi^+ \pi^-\pi^-$, $\pi^+\pi^-\pi^0\pi^0$,
$K^+K^-\pi^0\pi^0$, $K_SK_S\pi^+\pi^-$, $K_SK_S\pi^0\pi^0$,
$K_LK_L \pi^+\pi^-$, $K_LK_L\pi^0\pi^0$, $\eta\pi^0K^+K^-$,
$\eta\pi^0 K_SK_S$, $\eta\pi^+ K^- K^0$, $\eta \pi^- K^+ {\bar
{K^0}}$, $\eta \pi^0K_LK_L$. There is also the possible final
state $\pi^0\pi^0\pi^0\pi^0$ with ${\cal L}=9$.

The absence of strange quarks inside this $p\bar p$ bound state
may render the partial width of $K\bar K \pi$ channels smaller
than that of $\eta \pi\pi$  if the decay happens through the
regrouping of the quark and antiquarks inside the proton and
antiproton instead of through quark antiquark annihilation. If
selection rule allows, the three pairs of light quark and
anti-quark may recombine with each other to form three color
singlets easily, which eventually decay into three mesons. One, or
two or all of the resulting three mesons may decay into lighter
meson final states like multiple pions if symmetry allows.

OBELIX Collaboration observed that $p{\bar p}\to K^+K^-$ cross
section is much smaller than that of $p{\bar p}\to\pi^+\pi^-$ at
very low momenta around $50$ MeV \cite{lear}. In fact the total
$p\bar p$ annihilation cross section is dominated by the multipion
channels like $3\pi, 4\pi, 5\pi$. Especially the cross section of
the $5\pi$ channel is three times larger than that of $3\pi, 4\pi$
channels at $43.6$ MeV \cite{lear}.

We eagerly call for BES experimentalists to make strong efforts to
search for this very interesting state in the above listed final
states, especially with smaller ${\cal L}$. Confirmation of the
existence of this state in the meson final states will help unveil
the mysterious underlying dynamics of this multiquark system, thus
deepen our knowledge of the low energy sector of QCD.

Finally, we note in passing that although BES measurement of the
photon polar angle distribution favors $J=0$, the possibility of
$J=1$ is not absolutely excluded yet. If this state has
$J^{PC}I^G=1^{++}1^-$, it can decay into $(\rho
\pi)_{\mbox{S-wave}}$, $(\rho \pi)_{\mbox{D-wave}}$, $f_0(980)\pi$
etc. The absence of clear signal in the three-pion final states
has excluded such quantum number assignment.

With $J^{PC}I^G=1^{++}0^+$, the possible final states with the
lowest total partial wave number ${\cal L}=1$ are $K^+K^-\pi^0$,
$K_LK_L\pi^0$, $K_SK_S\pi^0$, $K^+{\bar {K^0}}\pi^-$,
$K^-K^0\pi^+$, $\eta \pi^0\pi^0$, $\eta K^+K^-$, $\eta K_SK_S$,
$\eta K_LK_L$. We do not list the possible final states with
${\cal L}=2, 4, 6$ here. Interested readers may consult Ref.
\cite{gao} or PDG \cite{pdg}.

In the extreme case that no narrow resonance with
$J^{PC}I^G=0^{-+}0^+$ is observed in the above mesonic decay
channels, it is worthwhile making some efforts to look for
possible $p\bar p$ resonances with $J^{PC}I^G=1^{++}0^+$ in the
mesonic final states.

S.L.Z. thanks Professors Y.-B. Dai, B.-S. Zou, W.-Z. Deng and
Z.-X. Zhang for useful discussions. C.S.G was supported by
National Natural Science Foundation of China under Grant No.
90103019, the Doctoral Program Foundation of Institution of Higher
Education, and the State Education Commission of China under Grant
No. 2000000147. S.L.Z was supported by the National Natural
Science Foundation of China, Ministry of Education of China,
FANEDD and SRF for ROCS, SEM.


\end{document}